\documentclass[article,preprint,groupedaddress]{revtex4}
\usepackage{epsfig}
\usepackage{graphicx}
\usepackage{amssymb}

\begin{document}
\title{Dragging of inertial frames in the composed Kerr-Newman-orbiting-ring system}
\author{Shahar Hod}
\address{The Ruppin Academic Center, Emeq Hefer 40250, Israel}
\address{ }
\address{The Hadassah Institute, Jerusalem 91010, Israel}
\date{\today}

\begin{abstract}
\ \ \ The dragging of inertial frames by an orbiting object, a well known phenomenon 
in general relativity, implies that the horizon angular velocity $\Omega^{\text{BH-ring}}_{\text{H}}$ 
of a central black hole in a composed black-hole-orbiting-ring system is no longer related to its 
angular-momentum $J_{\text{H}}$ by the familiar vacuum functional relation
$\Omega_{\text{H}}(J_{\text{H}})=J_{\text{H}}/M\alpha$ 
(here $\{M,\alpha\}$ are respectively the mass and normalized area of the central spinning black hole). 
Using a continuity argument, according to which the black-hole angular velocity changes smoothly 
during an adiabatic assimilation process of an orbiting ring, it has recently been revealed that 
the composed Kerr-ring system is characterized by the universal (that is, spin-{\it independent}) relation 
$\Delta\Omega_{\text{H}}\equiv\Omega^{\text{BH-ring}}_{\text{H}}(J_{\text{H}},J_{\text{R}},R\to
R^{+}_{\text{H}})-\Omega^{\text{Kerr}}_{\text{H}}(J_{\text{H}})={{J_{\text{R}}}/{4M^3}}$, 
where $\{R,J_{\text{R}}\}$ are respectively the radius of the ring and its orbital angular momentum 
and $R_{\text{H}}$ is the horizon radius of the central Kerr black hole. 
This intriguing observation naturally raises the following physically interesting question: Does the physical 
quantity $\Delta\Omega_{\text{H}}$ in a composed black-hole-orbiting-ring system is always characterized by the 
near-horizon functional relation $\Delta\Omega_{\text{H}}={{J_{\text{R}}}/{4M^3}}$ which 
is independent of the spin (angular momentum) $J_{\text{H}}$ of the central black hole? 
In order to address this important question, in the present compact paper 
we explore the physical phenomenon of dragging of inertial frames by an orbiting ring in the 
composed Kerr-Newman-black-hole-orbiting-ring system. In particular, using analytical techniques, 
we reveal the fact that in this composed two-body (black-hole-ring) system 
the quantity $\Delta\Omega_{\text{H}}$ has an explicit non-trivial functional dependence 
on the angular momentum $J_{\text{H}}$ of the central spinning black hole.  
\end{abstract}
\bigskip
\maketitle


\section{Introduction}

The gravitational two-body system has attracted the attention of 
physicists and mathematicians since the early days of general relativity. 
In particular, dynamical models which are based on the non-linearly coupled Einstein-matter field equations 
suggest that orbiting rings may form around central black holes during 
astrophysically realistic collapse scenarios of spinning self-gravitating compact stars \cite{Ans1,Shap,Shib}. 
Likewise, the coalescence process of two orbiting objects may produce a non-vacuum curved spacetime which is composed of a 
central spinning black hole that is gravitationally coupled to an external ring of orbiting matter \cite{Ans1,Shap,Shib}. 

It is therefore of physical interest to explore the physical and mathematical properties of the 
gravitationally-bound composed black-hole-orbiting-ring system. 
In particular, the influence of the external orbiting ring on the physical properties 
of the central black hole has been analyzed in \cite{Ans1,Shap,Shib,Will1,Will2}. 
Interestingly, using a perturbative scheme, Will \cite{Will1,Will2} (see also \cite{Hod1}) 
has studied analytically the physical properties of the composed 
black-hole-ring system in the small rotation limit of the central black hole. 

Intriguingly, it has been established in 
\cite{Will1,Will2} that, due to the general relativistic phenomenon of frame dragging which is caused by the 
presence of the orbiting ring, the familiar linear functional relation
\begin{equation}\label{Eq1}
\Omega^{\text{vacuum}}_{\text{H}}(J_{\text{H}})={{J_{\text{H}}}\over{M\alpha}}\
\end{equation}
between the horizon angular velocity $\Omega_{\text{H}}$ and the black-hole 
angular momentum $J_{\text{H}}$ is no longer valid in the composed two-body (black-hole-ring) system 
[here $\{M,\alpha\equiv A/4\pi\}$ 
are respectively the mass and the normalized surface area of the central black hole and 
$\Omega_{\text{H}}$ is the horizon angular-velocity of the central black hole]. 

In particular, it has been established \cite{Will1,Will2} that, due to the dragging of inertial frames 
by the orbiting ring, a central black hole with zero angular momentum in 
the composed black-hole-orbiting-ring system may be characterized by a non-zero angular-velocity
\cite{Notecor1}
\begin{equation}\label{Eq2}
\Omega^{\text{BH-ring}}_{\text{H}}(J_{\text{H}}=0,J_{\text{R}},R)={{2J_{\text{R}}}\over{R^3}}\
,
\end{equation}
where $\{R,J_{\text{R}}\}$ are respectively the proper circumferential radius and the angular-momentum 
of the orbiting ring. 

The slowly rotating black holes that were studied analytically in \cite{Will1,Will2} are 
characterized by the simple leading-order mass-radius relation $R^{+}_{\text{H}}=2M$ (here $R^{+}_{\text{H}}$ is the radius of the 
black-hole outer horizon), which yields the asymptotic functional behavior 
\begin{equation}\label{Eq3}
\Omega^{\text{BH-ring}}_{\text{H}}(J_{\text{H}}=0,J_{\text{R}},R\to
R^{+}_{\text{H}})\to {{J_{\text{R}}}\over{4M^3}}\
\end{equation}
for the angular velocity of the central black hole at the assimilation point $R\to R^{+}_{\text{H}}$ of the orbiting ring.  

Intriguingly, we have recently revealed the fact \cite{HodKerr} that the near-horizon limiting functional behavior (\ref{Eq3}) for 
the perturbed angular velocity $\Omega^{\text{BH-ring}}_{\text{H}}(J_{\text{H}},J_{\text{R}},R\to
R^{+}_{\text{H}})$ of the central black hole just before 
it assimilates the orbiting ring can be obtained using a {\it continuity} argument which is 
based on the {\it smooth} (continuous) evolution of the black-hole angular velocity 
during an adiabatic assimilation process of the ring into the central black hole. 

In particular, the physically motivated continuity argument states that, in an {\it adiabatic} physical process 
in which the ring is slowly lowered into a central black hole, 
the angular velocity $\Omega^{\text{BH-ring}}_{\text{H}}(J_{\text{H}}=0,J_{\text{R}},R\to
R^{+}_{\text{H}})$ of the perturbed black hole at the 
assimilation point $R\to R^{+}_{\text{H}}$ of the ring (that is, just {\it before} the black-hole-ring merger) is 
equal to the angular velocity $\Omega^{\text{Kerr}}_{\text{H}}$ of the final Kerr black-hole configuration 
(that is, the bare black hole that forms {\it after} the assimilation of the ring) whose total 
angular momentum is given, due to angular momentum conservation, 
by the relation $J^{\text{new}}_{\text{H}}=J_{\text{H}}+J_{\text{R}}$ . 

Using the continuity relation
\begin{equation}\label{Eq4}
\Omega^{\text{BH-ring}}_{\text{H}}(J_{\text{H}},J_{\text{R}},R\to
R^{+}_{\text{H}})=\Omega^{\text{Kerr}}_{\text{H}}(J^{\text{new}}_{\text{H}}=J_{\text{H}}+J_{\text{R}})\
\end{equation}
for the composed Kerr-black-hole-orbiting-ring system, we have recently derived the remarkably 
compact leading-order (that is, linear in the angular momentum $J_{\text{R}}$ of the orbiting ring) 
functional relation \cite{HodKerr}
\begin{equation}\label{Eq5}
\Delta\Omega_{\text{H}}(R\to
R^{+}_{\text{H}})={{J_{\text{R}}}\over{4M^3}}\  ,
\end{equation}
where 
\begin{equation}\label{Eq6}
\Delta\Omega_{\text{H}}(R\to
R^{+}_{\text{H}})\equiv\Omega^{\text{BH-ring}}_{\text{H}}(J_{\text{H}},J_{\text{R}},R\to
R^{+}_{\text{H}})-\Omega^{\text{Kerr}}_{\text{H}}(J_{\text{H}})
\end{equation}
is the asymptotic (near merger) deviation of the angular velocity
$\Omega^{\text{BH-ring}}_{\text{H}}(J_{\text{H}})$ of the perturbed black hole in the composed
black-hole-orbiting-ring system from the familiar angular velocity
$\Omega^{\text{Kerr}}_{\text{H}}(J_{\text{H}})={{J_{\text{H}}}/{M\alpha}}$ of a bare (vacuum) Kerr black hole with the
same angular-momentum $J_{\text{H}}$. 

Interestingly, it has been pointed out in \cite{HodKerr} that the analytically derived functional expression (\ref{Eq5}) 
for the angular velocity deviation function $\Delta\Omega_{\text{H}}$ is {\it universal} for all 
Kerr black holes in the sense that it is {\it independent} of the 
angular momentum $J_{\text{H}}$ of the central spinning black hole. 
This physically intriguing observation naturally raises the following important question: 
Does the physical quantity $\Delta\Omega_{\text{H}}$ in a composed black-hole-orbiting-ring system is always characterized by the 
functional relation $\Delta\Omega_{\text{H}}={{J_{\text{R}}}/{4M^3}}$ which, as emphasized above, 
is independent of the central black-hole angular momentum $J_{\text{H}}$? 

In order to address this physically intriguing question, in the present compact paper we shall explore, using 
analytical techniques, the phenomenon of frame dragging by an orbiting ring in 
a composed Kerr-Newman-black-hole-orbiting-ring system. 
Our analysis, to be presented below, reveals the fact that the physical quantity $\Delta\Omega_{\text{H}}$ 
in the composed Kerr-Newman-black-hole-ring system is non-universal. 
In particular, we shall explicitly prove that the 
angular velocity deviation function $\Delta\Omega_{\text{H}}(R\to
R^{+}_{\text{H}})$ has an explicit functional dependence 
on the angular momentum $J_{\text{H}}$ of the central (spinning and charged) black hole.  

\section{The angular-velocity/angular-momentum functional relation in the composed Kerr-Newman-black-hole-orbiting-ring system}

In the present section we shall analyze the functional dependence 
$\Omega^{\text{BH-ring}}_{\text{H}}=\Omega^{\text{BH-ring}}_{\text{H}}(J_{\text{H}},J_{\text{R}},R\to
R^{+}_{\text{H}})$ of the black-hole angular velocity on the angular momentum 
of an orbiting ring in the composed Kerr-Newman-black-hole-orbiting-ring system, 
where $R^{+}_{\text{H}}=M+(M^2-Q^2-a^2)^{1/2}$ is the outer horizon radius of a Kerr-Newman black hole of mass $M$, electric charge $Q$, and 
angular momentum $J_{\text{H}}\equiv Ma$. 

To this end, we shall use the physically motivated continuity argument presented in \cite{HodKerr} which 
implies the smooth functional relation
\begin{equation}\label{Eq7}
\Omega^{\text{BH-ring}}_{\text{H}}(J_{\text{H}},J_{\text{R}},R\to
R^{+}_{\text{H}})=\Omega^{\text{KN}}_{\text{H}}(J^{\text{new}}_{\text{H}}=J_{\text{H}}+J_{\text{R}})\
\end{equation}
at the capture point of the ring by the central spinning Kerr-Newman black hole in an adiabatic assimilation 
process. 
It is worth stressing the fact that the left-hand-side of (\ref{Eq7}) is the 
near-horizon $R\to R^{+}_{\text{H}}$ limit of the angular velocity which characterizes the 
central spinning black hole in the composed black-hole-orbiting-ring system just {\it before} the assimilation of
the ring into the black hole, whereas the right-hand-side of the continuity relation (\ref{Eq7}) is 
the angular velocity that characterizes the final spinning Kerr-Newman black hole {\it after} the absorption process.

The adiabatic assimilation process of the orbiting ring into the central black hole produces
a final spinning and charged Kerr-Newman black-hole spacetime which is characterized by 
the following modified physical parameters:
\begin{equation}\label{Eq8}
M\to M^{\text{new}}=M+{\cal E}({\text{R}_{\text{H}}})\ \ \ \ ; \ \ \ \ 
J_{\text{H}}\to J^{\text{new}}_{\text{H}}=J_{\text{H}}+J_{\text{R}}\ \ \ \ ; \ \ \ \ Q\to Q\  ,
\end{equation}
where $J_{\text{R}}$ is the angular momentum of the ring and ${\cal E}({\text{R}_{\text{H}}})$, 
the energy of the ring at the absorption point $R\to R^{+}_{\text{H}}$, is given by the leading-order (linear in $J_{\text{R}}$) 
functional expression \cite{Car,Bek1,Notered,Notecor2}
\begin{equation}\label{Eq9}
{\cal E}(R\to
R^{+}_{\text{H}})={{J_{\text{H}}}\over{M\alpha}}\cdot J_{\text{R}}\cdot[1+O(J_{\text{R}}/M^2)]\ ,
\end{equation}
where
\begin{equation}\label{Eq10}
\alpha=2MR_{\text{H}}-Q^2\
\end{equation}
is the normalized surface area of the central Kerr-Newman black hole. 

Before proceeding, it is worth emphasizing that we consider an idealized adiabatic assimilation process of
the orbiting ring into the central spinning black hole. 
In particular, following \cite{Bek1} we assume that the ring is slowly lowered towards the central black hole and it is 
therefore characterized by an almost zero radial momentum [see Eq. (\ref{Eq9})] at the assimilation point $R\to R^{+}_{\text{H}}$. 
Taking cognizance of Eqs. (\ref{Eq8}), (\ref{Eq9}), and (\ref{Eq10}) one finds that in this adiabatic process 
the surface area of the central black hole remains unchanged to linear-order in the angular momentum $J_{\text{R}}$ of the ring \cite{Notead}:
\begin{equation}\label{Eq11}
\alpha\to\alpha^{\text{new}}=\alpha\cdot[1+O(J^2_{\text{R}}/M^4)]\  .
\end{equation}

Substituting Eqs. (\ref{Eq8}), (\ref{Eq9}), and (\ref{Eq11}) into Eq. (\ref{Eq1}), one obtains the leading-order 
functional relation 
\begin{equation}\label{Eq12}
\Omega^{\text{KN}}_{\text{H}}(J^{\text{new}}_{\text{H}}=
J_{\text{H}}+J_{\text{R}})={{J_{\text{H}}}\over{M\alpha}}+
{{{\text{R}^2_{\text{H}}}}\over{M\alpha^2}}\cdot J_{\text{R}}\cdot[1+O(J_{\text{R}}/M^2)]\
\end{equation}
for the angular velocity of the final spinning and charged Kerr-Newman black hole {\it after} 
the adiabatic absorption of the orbiting ring.

Taking cognizance of Eqs. (\ref{Eq1}), (\ref{Eq7}), and (\ref{Eq12}), 
one obtains the characteristic leading-order angular-velocity/angular-momentum functional relation
\begin{equation}\label{Eq13}
\Omega^{\text{BH-ring}}_{\text{H}}(J_{\text{H}},J_{\text{R}},R\to
R^{+}_{\text{H}})=\Omega^{\text{KN}}_{\text{H}}(J_{\text{H}})+
{{{\text{R}^2_{\text{H}}}}\over{M\alpha^2}}\cdot J_{\text{R}}\
\end{equation}
for a central spinning black hole of intrinsic angular momentum $J_{\text{H}}$ 
in the composed Kerr-Newman-black-hole-orbiting-ring system, 
where $\Omega^{\text{KN}}_{\text{H}}(J_{\text{H}})=J_{\text{H}}/M\alpha$ [see Eq. (\ref{Eq1})] 
is the angular-velocity of a bare (ring-less) spinning Kerr-Newman 
black hole with the {\it same} value $J_{\text{H}}$ of intrinsic angular momentum. 
From the analytically derived relation (\ref{Eq13}) one finds the characteristic near-horizon 
functional behavior 
\begin{equation}\label{Eq14}
\Delta\Omega_{\text{H}}(R\to
R^{+}_{\text{H}})\equiv\Omega^{\text{BH-ring}}_{\text{H}}(J_{\text{H}},J_{\text{R}},R\to
R^{+}_{\text{H}})-\Omega^{\text{KN}}_{\text{H}}(J_{\text{H}})=
{{{\text{R}^2_{\text{H}}}}\over{M\alpha^2}}\cdot J_{\text{R}}\
\end{equation}
for the angular velocity deviation function. 

\section{Summary and discussion}

The dragging of inertial frames by an orbiting object is one of the most intriguing predictions of 
general relativity. In particular, Will \cite{Will1,Will2} has analyzed, using a perturbation scheme, the physical 
properties of the composed black-hole-orbiting-ring system in the slow rotation regime of the central black hole. 
Interestingly, it has been proved \cite{Will1,Will2} that, due to the dragging of inertial frames by the orbiting ring, a 
central black hole of zero angular momentum can have a non-trivial (non-zero) angular velocity which, to linear order in the 
angular momentum $J_{\text{R}}$ of the ring, is given by the near-horizon dimensionless 
functional expression [see Eq. (\ref{Eq3})]
\begin{equation}\label{Eq15}
M\Omega^{\text{BH-ring}}_{\text{H}}(J_{\text{H}}=0,J_{\text{R}},R\to
R^{+}_{\text{H}})\to {{J_{\text{R}}}\over{4M^2}}\  .
\end{equation}
 
Using a physically motivated continuity argument, 
which is valid in an adiabatic assimilation process in which the ring is slowly lowered into the central black hole, we have recently \cite{HodKerr} generalized the result (\ref{Eq15}) of \cite{Will1,Will2} to the regime 
of central black holes with non-zero angular momenta. 
In particular, we have revealed the physically intriguing fact that the composed Kerr-black-hole-orbiting-ring system 
is characterized by the universal (spin-{\it independent}) dimensionless functional relation \cite{HodKerr}
\begin{equation}\label{Eq16}
M\cdot\Delta\Omega^{\text{Kerr}}_{\text{H}}(R\to
R^{+}_{\text{H}})={{J_{\text{R}}}\over{4M^2}}\  ,
\end{equation}
where $\Delta\Omega_{\text{H}}(R\to
R^{+}_{\text{H}})\equiv\Omega^{\text{BH-ring}}_{\text{H}}(J_{\text{H}},J_{\text{R}},R\to
R^{+}_{\text{H}})-\Omega^{\text{Kerr}}_{\text{H}}(J_{\text{H}})$ is the asymptotic (near merger) deviation function 
of the angular velocity of the central spinning black hole in the composed
black-hole-orbiting-ring system from the familiar vacuum value 
$\Omega^{\text{Kerr}}_{\text{H}}(J_{\text{H}})={{J_{\text{H}}}/{M\alpha}}$ that characterizes 
the angular velocity of the bare (ring-less) Kerr black hole with the same angular momentum $J_{\text{H}}$. 

The question whether the angular velocity deviation function $\Delta\Omega_{\text{H}}(R\to
R^{+}_{\text{H}})$ always has a universal form which is independent of the angular momentum (spin) $J_{\text{H}}$ of the 
central black hole [as was shown to be the case in the composed Kerr-ring system \cite{HodKerr}, see Eq. (\ref{Eq5})] 
arises quite naturally. 
In the present compact paper we have studied this important issue by analyzing the functional behavior of 
the angular velocity deviation function $\Delta\Omega^{\text{KN}}_{\text{H}}(R\to
R^{+}_{\text{H}})$ that characterizes the composed Kerr-Newman-black-hole-orbiting-ring system. 
 
In particular, analyzing the physical phenomenon of frame dragging by an orbiting ring in the 
composed Kerr-Newman-ring system and using the continuity relation (\ref{Eq7}), 
which reflects a smooth functional behavior of the angular velocity of the central black hole during an adiabatic
assimilation process in which the ring is absorbed by the spinning black hole, 
we have derived the leading-order (linear in the angular momentum $J_{\text{R}}$ of the orbiting ring) 
spin-dependent dimensionless functional expression [see Eqs. (\ref{Eq10}) and (\ref{Eq14})] \cite{NoteKKN}
\begin{equation}\label{Eq17}
M\cdot\Delta\Omega^{\text{KN}}_{\text{H}}(R\to
R^{+}_{\text{H}})={{{\text{R}^2_{\text{H}}}}\over{[\text{R}^2_{\text{H}}+(J_{\text{H}}/M)^2]^2}}\cdot J_{\text{R}}\
\end{equation}
for the non-trivial angular velocity deviation function 
in the composed two-body (Kerr-Newman-ring) system. 

From the analytically derived functional relation (\ref{Eq17}) one learns that, in the 
composed Kerr-Newman-black-hole-orbiting-ring system, the angular velocity deviation function is non-universal. 
In particular, it has an explicit functional dependence on the spin $J_{\text{H}}$ of the central 
black hole \cite{NoteQ0}. 

\bigskip
\noindent
{\bf ACKNOWLEDGMENTS}
\bigskip

This research is supported by the Carmel Science Foundation. I thank
Yael Oren, Arbel M. Ongo, Ayelet B. Lata, and Alona B. Tea for
stimulating discussions.



\begin{thebibliography}{99}

\bibitem{Ans1} M. Ansorg and D. Petroff, Phys. Rev. D {\bf 72}, 024019 (2005).

\bibitem{Shap} N. L. Shapiro, Astrophys. J. {\bf 444}, 306 (1995).

\bibitem{Shib} M. Shibata, K. Taniguchi, and K. Uryu, Phys. Rev. D {\bf 68}, 084020 (2003).

\bibitem{Will1} C. M. Will, Astrophys. J. {\bf 191}, 521 (1974).

\bibitem{Will2} C. M. Will, Astrophys. J. {\bf 196}, 41 (1975).

\bibitem{Hod1} S. Hod, Phys. Rev. D {\bf 87}, 024036 (2013)
[arXiv:1311.1281]; S. Hod, Phys. Lett. B {\bf 726}, 533 (2013)
[arXiv:1312.4969]; S. Hod, The Euro. Phys. Jour. C {\bf 74}, 2840
(2014) [arXiv:1404.1566].

\bibitem{Notecor1} As explicitly proved in \cite{Will1,Will2}, 
the expression (\ref{Eq2}) for the angular velocity of the black-hole horizon is valid in the perturbative regime
$J_{\text{R}}/R^2\ll1$. Sub-leading corrections to the functional relation (\ref{Eq2}) are 
of order $O(J^2_{\text{R}}/R^5)$.

\bibitem{HodKerr} S. Hod, The Euro. Phys. Jour. C {\bf 75}, 541 (2015) [arXiv:1511.02964].

\bibitem{Car} B. Carter, Phys. Rev. {\bf 174}, 1559 (1968).

\bibitem{Bek1} J. D. Bekenstein, Phys. Rev. {\bf 7}, 2333 (1973).

\bibitem{Notered} Note that the mass-energy 
of the ring 
is completely red-shifted at the assimilation point $R\to R^{+}_{\text{H}}$. 

\bibitem{Notecor2} Note that the expression (\ref{Eq9}) for the energy of the ring at the assimilation point $R\to R^{+}_{\text{H}}$ 
is valid in the perturbative regime $J_{\text{R}}/R^2\ll1$. Sub-leading corrections to the functional relation (\ref{Eq9}) are 
of order $O(J^2_{\text{R}}/R^3,\mu^2/R)$.

\bibitem{Notead} As explicitly proved by Bekenstein \cite{Bek1}, the black-hole surface area $\alpha$ plays the
role of an adiabatic invariant.

\bibitem{NoteKKN} Note that the analytically derived functional expression (\ref{Eq17}) for the 
composed Kerr-Newman-black-hole-orbiting-ring system reduces to the simpler expression (\ref{Eq5}) of \cite{HodKerr} 
in the $Q\to0$ limit (in which case $\alpha=2MR^{+}_{\text{H}}$). 

\bibitem{NoteQ0} It is important to point out that the $Q=0$ case is characterized by the relation 
$R^{+}_{\text{H}}=M+[M^2-(J_{\text{H}}/M)^2]^{1/2}$ for the outer horizon radius of a neutral Kerr black hole. 
Substituting this relation into Eq. (\ref{Eq17}), one obtains the $J_{\text{H}}$-{\it independent} relation 
$\Delta\Omega_{\text{H}}(R\to R^{+}_{\text{H}})={{J_{\text{R}}}/{4M^3}}$ for Kerr 
black holes. As a consistency check we emphasize the fact that this result agrees with the analytical 
results presented in \cite{HodKerr} for neutral Kerr black holes. 

\end{thebibliography}
\end{document}